\RequirePackage{fix-cm}

\documentclass[natbib,smallextended]{svjour3}
\usepackage{amsmath}     
\usepackage{wasysym}           
\usepackage{graphicx}
\usepackage{amssymb}
\usepackage{epstopdf}
\usepackage{mathrsfs}
\usepackage{anyfontsize}
\usepackage{natbib}
\usepackage{color}
\usepackage{lipsum}
\usepackage{diagbox}
\smartqed

\begin{document}
\title{\protect The Influence of Black Hole Binarity on Tidal Disruption Events}
\subtitle{}
\author{Eric R.~Coughlin\thanks{Corresponding Author} \and 
Philip J.~Armitage \and 
Giuseppe Lodato \and 
C.~J.~Nixon
}

\institute{Eric R.~Coughlin\thanks{Corresponding Author} \at 
Department of Astrophysical Sciences, Princeton University, Princeton, NJ 08544, USA \\
Columbia Astrophysics Laboratory, Columbia University, New York, NY 10027, USA \\ 
\email{eric.r.coughlin@gmail.com}
\and
Philip J.~Armitage \at 
Department of Physics and Astronomy, Stony Brook University, Stony Brook, NY 11790, USA \\
Center for Computational Astrophysics, Flatiron Institute, 162 Fifth Avenue, New York, NY 10010, USA \\ 
JILA, University of Colorado \& NIST, 440 UCB, Boulder, CO 80309, USA
\and
Giuseppe Lodato \at 
Dipartimento di Fisica, Universit\`a degli Studi di Milano, Via Celoria 16, 20133 Milano, Italy 
\and
C.~J.~Nixon \at
Department of Physics and Astronomy, University of Leicester, Leicester, LE1 7RH, UK
}

\date{Received: date / Accepted: date}

\maketitle

\begin{abstract}
Mergers are fundamental to the standard paradigm of galaxy evolution, and provide a natural formation mechanism for supermassive black hole binaries. The formation process of such a binary can have a direct impact on the rate at which stars are tidally disrupted by one or the other black hole, and the luminous signature of the tidal disruption itself can have distinct imprints of a binary companion. In this chapter we review our current understanding of the influence of black hole binarity on the properties of tidal disruption events. We discuss the rates of tidal disruption by supermassive black hole binaries, the impact of a second black hole on the fallback of debris and the formation of an accretion flow, and the prospects for detection of tidal disruption events by supermassive black hole binaries. 
\end{abstract}


\section{Introduction}
Supermassive black hole (SMBH) binaries are a byproduct of the hierarchical picture of galactic evolution. When two galaxies merge, the constituent black holes inspiral over the course of Myrs {-- Gyrs (see Equation \ref{eq:tfric})} through dynamical friction \citep{chandrasekhar60}, whereby the ``wake'' of stars generated by the gravitational focusing of the black holes saps their kinetic energy. This process continues to reduce the separation between the SMBHs until the SMBH binary circular velocity exceeds the stellar velocity dispersion of the stars within the merged galaxy, at which point {it becomes a bound system in the sense that the two SMBHs feel each other's gravitational presence. The binary} becomes ``hard'' and only individual stellar interactions -- which eject stars from the galaxy at high velocities -- continue to shrink the binary separation {once it reaches a critical separation $a_{\rm h} \sim GM_2/4\sigma^2$, where $M_2$ is the mass of the secondary and $\sigma$ is the velocity dispersion} (e.g., \citealt{begelman80, quinlan96, merritt01, sesana06, kelley17}). Depending on the specific properties of the SMBH binary and the structure of the galactic nucleus, the binary can stall at parsec-scale separations for Gyr before gravitational radiation results in their coalescence, though gas dynamics (e.g., \citealt{dunhill14,goicovic16}) and third bodies (e.g., \citealt{ryu18, bonetti19}) may alleviate this ``last parsec problem.''

Theoretical predictions for how SMBH binaries form and coalesce within merged galaxies need to be tested across a large range of spatial separations, where different physical processes dominate. On kpc-scales there is a growing body of informative observational data from the detection of dual Active Galactic Nuclei (AGN) (e.g., \citealt{comerford13}). In this Chapter we argue that observing the tidal disruption of stars by one black hole in a binary system -- on scales roughly a million times smaller than dual AGN -- could yield valuable constraints on merger physics immediately prior to {(i.e., when the SMBH binary is within the gravitational-wave inspiral regime)} the final coalescence detectable in gravitational waves. However, realizing this opportunity to probe the SMBH binary population on small scales is only achievable once the effects of binarity on the evolution of tidal disruption events (TDEs) are understood from theoretical grounds. 

Here we discuss theoretical progress toward understanding the impact of black hole binarity on the properties of TDEs. In Section \ref{sec:elementary}, we outline some very basic arguments concerning the SMBH binary separations and mass ratios at which we expect certain physical mechanisms to alter the rate and appearance of TDEs. In Section \ref{sec:rates}, we discuss the recent work that has been done on the rates of TDEs by black holes in binaries, and Section \ref{sec:fallback} describes the influence of a binary companion on the fallback and accretion of debris following a TDE. Section \ref{sec:prospects} gives a discussion of the prospects of detecting TDEs from binaries, and we summarize in Section \ref{sec:summary}.

\section{Elementary considerations}
\label{sec:elementary}
The qualitative evolution of supermassive black hole binaries that form following galactic mergers follows the scenario described by \cite{begelman80}. There are three distinct phases.

Immediately following the merger, the two black holes orbit in the gravitational potential of the merged galaxy, with a separation that may initially be of the order of a few kpc. Dynamical friction against stars (and gas, if present) causes the separation to shrink. If the galaxy has a core containing $N$ stars within radius $r$, with one-dimensional velocity dispersion $\sigma$, the time scale for orbital decay for a black hole of mass M is\footnote{We note, however, that at early times each black hole ``sees'' the other galactic nucleus as a single entity with a mass much greater than its mass, which enhances the inspiral rate at early times.} \citep{yu02},
\begin{equation}
    t_{\rm fric} \sim \frac{4 \times 10^6}{\log N} 
        \left( \frac{\sigma}{200 \ {\rm km \ s^{-1}}} \right) 
        \left( \frac{r}{100 \ {\rm pc}} \right)^2
        \left( \frac{M}{10^8 \ M_\odot} \right)^{-1} \ {\rm yr}. \label{eq:tfric}
\end{equation}
Dynamical friction is efficient enough to bring binaries to separations of the order of $a \sim 10-100 \ {\rm pc}$ (depending on the galaxy), {\em unless} the secondary is of very low mass. Very low mass secondaries formed from mergers will instead continue to orbit with minimal energy loss in the merged galaxy.

A bound binary forms at $a \sim \sqrt{GM/\sigma^2}$, when the binary separation matches the sphere of influence for the total mass $M$. Further orbital decay is accompanied by an increase in the orbital velocity and reduced effectiveness of dynamical friction. By the time the binary passes the threshold separation to be hard \citep[in the sense of][]{heggie75,binney87,quinlan96,yu02},
\begin{equation}
    a_{\rm h} = \frac{GM_2}{4 \sigma^2}, \label{ah}
\end{equation}
the dominant energy loss process is not dynamical friction but rather the scattering of individual stars. Scattering by a hard black hole binary generally removes the intruding stars from the vicinity of the binary, {meaning that stars that possess pericenters within a few binary separations belong to a {\em loss cone} within the stellar energy-angular momentum phase space distribution; if the refilling of this cone by two-body relaxation does not occur quickly enough, this region of the phase space is empty} \citep{frank76}. The rate of decay is then determined by whatever processes bring fresh stars into dynamical contact with the binary \citep{alexander17}. For spherical or axisymmetric stellar systems in massive galaxies the maximum predicted inspiral time can exceed the Hubble time \citep{yu02}, leading to the prediction that some binaries might stall at roughly 0.1-1~pc scales \citep[the ``last parsec problem'';][]{begelman80}. There is no last parsec problem in triaxial galactic nuclei \citep{berczik06,holley-bockelmann06,gualandris17}, or in general for low mass black holes in cuspy stellar distributions. Angular momentum can also be exchanged with gas disks if there is sufficient gas present on sub-pc scales  \citep{armitage02,dotti07,cuadra09,lodato09,nixon11,tang17}. 

Finally, energy loss via gravitational waves controls the final phase of inspiral. At fixed mass ratio and eccentricity the merger time due to gravitational waves scales as $t_{\rm gr} \propto a^4 / M^3$, so there is a sharp transition between the stellar (or gas) dynamical phase of merger and that dominated by gravitational radiation. 

\begin{figure*}
\centering
\includegraphics[width=0.495\textwidth]{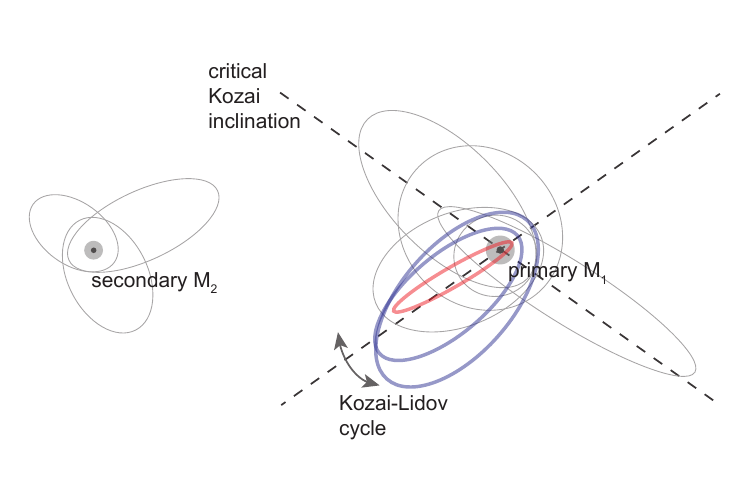}
\includegraphics[width=0.495\textwidth]{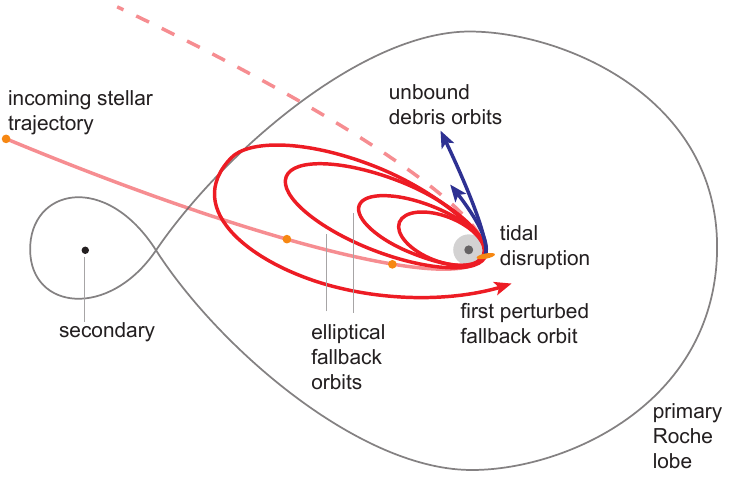}
\caption{Illustration of two of the ways by which supermassive black hole binaries impact the physics of tidal disruption events. At relatively large binary separations ({\em left}) the gravitational potential of the companion can excite Kozai-Lidov oscillations among bound stars whose orbits are inclined at $i > i_{\rm crit}$. The cycles increase the orbital eccentricity at fixed semi-major axis, potentially reducing the pericenter distance to within the tidal radius. This can increase the {\em rate} of TDEs. At typically smaller binary separations ({\em right}) the binary potential can affect the {\em dynamics} of the debris stream. Fallback orbits whose apocenter distance approaches the size of the disrupting black hole's Roche lobe will be significantly perturbed, modifying the fallback rate.}
\label{fig:figure_BBH_cartoon}
\end{figure*}

Depending upon how far a binary has advanced through the above sequence we can identify distinct effects that binarity may have on TDEs. For ``wide" binaries---where wide here typically means any separation greater than $\sim 10^{-3} \, {\rm pc}$---neither the dynamics of the disruption nor that of the fallback is appreciably altered by the presence of the second black hole. The presence of a binary may be inferred indirectly, either by observing TDEs that are displaced from the center of galaxies, or by measuring a different {\em rate} of TDEs in post-merger galaxies compared to similar galaxies that have not recently undergone merger. In galaxies with only one black hole, {the majority of} tidally disrupted stars originate from {the boundary between the full and empty loss cone, which corresponds roughly to} the sphere of influence \citep{alexander03, stone16}, so at a minimum we expect that changes to the rate of TDEs will occur around the time that a bound supermassive black hole binary forms. The rate can be altered either by changes in how unbound stars are scattered into the loss cones of the black holes, or by dynamics that the binary companion induces in the bound stars surrounding either hole. Kozai-Lidov oscillations, illustrated in Figure~\ref{fig:figure_BBH_cartoon}, are one binary specific process that can affect the rate of bound star disruptions. 
Galactic mergers may also lead to the formation of new populations of stars, for example in eccentric disks, whose dynamics can have a large impact on TDE rates \citep{madigan18}. {If the rates are increased dramatically, a single galaxy may also exhibit multiple flares within $\sim$ year timescales, which would also be a good, indirect indicator of the presence of a SMBH binary \citep{wegg11,thorp18}.}

For close binaries, the dynamics of the debris stream that provides the fallback material can be perturbed by the gravitational influence of the other black hole. Let us sketch out the simplest estimate of when such perturbations have an observable effect on TDEs (via, e.g., altering the fallback rate, initiating accretion on to {\em both} black holes, or changing how streams collide and circularize). We consider a binary of total mass $M$, made up of two equal mass black holes with $M_1 = M_2 = M/2$. The binary is on a circular orbit, with separation $a \gg r_{\rm t}$, the tidal disruption radius. After a star has been tidally disrupted by one of the holes, the debris stream follows a set of highly eccentric orbits with $e \simeq 1$. At time $T$ after the disruption, the relation between the time and the apocenter of the debris that is only then returning to the vicinity of the black hole is,
\begin{equation}
 T^2 = \frac{\pi^2 a_{\rm apo}^3}{GM}.
\end{equation}
Clearly, the fallback dynamics will be drastically perturbed if $a_{\rm apo} > a/2$. (For general mass ratios, as illustrated in Figure~\ref{fig:figure_BBH_cartoon}, strong perturbations occur if the debris, at apocenter, leaves the Roche lobe of the disrupting black hole.) Imposing this condition we find that fallback in a binary system differs strongly from the single black hole case, at time $T$, if the binary has separation,
\begin{eqnarray}
 a & \leq & \left( GM \right)^{1/3} P^{2/3}, \nonumber \\
   & = & 3 \times 10^{-4} \left( \frac{M}{10^6 \ M_\odot} \right)^{1/3} 
   \left( \frac{T}{1 \ {\rm month}} \right)^{2/3} \ {\rm pc}.
\label{eq:a_perturb}   
\end{eqnarray}
(Note that we have dropped unimportant numerical factors from this estimate.) This separation is interestingly small. If we observe TDEs whose dynamics is directly influenced by the binary potential, we will be probing an unexplored regime of supermassive black hole binaries much tighter than any securely identified to date.

At the small separations where we expect perturbations to the fallback to be observable, the binary will be decaying due to gravitational wave emission. This imposes a strong bias toward seeing binary-modified TDEs around the lowest mass black holes. Evaluating the gravitational wave merger time at the critical separation given by equation~(\ref{eq:a_perturb}) we find,
\begin{eqnarray}
 t_{\rm merge} & \sim & \frac{5 c^5}{64 G^{5/3}} M^{-5/3} T^{8/3}, \nonumber \\
 & \sim & 2 \times 10^7 \left( \frac{M}{10^6 \ M_\odot} \right)^{-5/3}
 \left( \frac{T}{1 \ {\rm month}} \right)^{8/3} \ {\rm yr}.
\end{eqnarray}
The persistence of low mass binaries close enough to appreciably affect TDE fallback dynamics is much greater than for higher mass systems. We therefore expect there to be a strong bias toward seeing ``binary TDEs" around low mass supermassive black holes, {\em in addition to} the usual bias toward low mass black holes that arise even for normal TDEs simply because low mass systems are more numerous. {The peak fallback rate $\dot{M}$ onto the black hole is also expected to scale with the black hole mass $M$ as $\dot{M} \sim M^{-1/2}$ \citep{rees88}, making TDEs by larger black holes less luminous overall and harder to detect, though we note that the more relativistic tidal radii of larger black holes enhances general relativistic precession and should facilitate disc formation.}

Finally, when $a \sim r_{\rm t}$ the physics of the disruption itself will be sensitive to the presence of a binary potential. The merger time scale at this separation is very short, so absent some special process that synchronizes disruption with merger it is unlikely that TDEs will be observed in this regime.

\section{TDE rates in black hole binary systems}
\label{sec:rates}
When the galactic merger that initiates the formation of the SMBH binary is still in its early stages and the SMBHs are at kpc separations, the TDE rate -- onto one or the other black hole -- is likely not greatly enhanced above the typical value that results from loss cone refilling (though triaxiality and chaotic stellar orbits induced by the merger could conceivably enhance the rate; \citealt{merritt04}). Once the binary separation becomes comparable to the sphere of influence of the primary, the secondary encounters the Bahcall-Wolf ``cusp'' around the primary \citep{bahcall76}, and the presence of the secondary can start to significantly modify the stellar dynamics and, correspondingly, the rate at which stars are tidally disrupted.

As mentioned in Section \ref{sec:elementary}, one modification to the TDE rate arises from Kozai-Lidov oscillations \citep{lidov62,kozai62}, whereby the secondary causes circumprimary stars on highly-inclined orbits (relative to the binary orbital plane) to be driven to low-angular momentum orbits and disrupted by the primary; see the left panel of Figure \ref{fig:figure_BBH_cartoon}. This stage of the problem was analyzed in detail by \citet{ivanov05}, who demonstrated analytically that the rate of TDEs -- exclusively by the primary -- can be increased to $\dot{N}_{\rm TDE} \simeq 0.1 - 1$ gal$^{-1}$yr$^{-1}$ when the black hole mass ratio is fairly small (such that one can approximate the gravitational potential as being dominated by the primary for those stars undergoing Kozai oscillations and contributing to the increased rate). The timescale over which the orbits of stars are altered by the Kozai process, and thereby the time over which the rate of TDEs is expected to be enhanced by this process, is \citep{ivanov05}

\begin{equation}
    T_{\rm K} \simeq \frac{1}{q}\left(\frac{a}{D}\right)^{-3/2}\sqrt{\frac{D^3}{GM}},
\end{equation}
where $q$ is the binary mass ratio, $a$ is the distance of a star from the primary, $D$ is the binary separation, and $M$ is the mass of the primary (assumed to be much more massive than the secondary). Taking $D = 1$ pc, $a/D = 0.5$, $M = 10^{7} M_{\odot}$, and $q = 0.1$, this timescale is $T_{\rm K} \simeq 10^{5}$ yr. Therefore, the enhancement to the TDE rate resulting from Kozai oscillations induced by the secondary is very short-lived.

An additional modification to the TDE rate arises from direct interactions between the secondary and individual, bound stars. \citet{chen09} demonstrated that these interactions generate highly complex, restricted three-body orbits as stars are scattered off of the secondary, which -- similar to the findings of \citet{ivanov05} -- generates a short-lived ($\lesssim 10^{6}$ yr) burst of TDEs. However, in contrast to the mechanism proposed by \citet{ivanov05}, which relied on Kozai-Lidov oscillations to fill the loss cone of the primary, \citet{chen09} found that the majority of disrupted stars entered the loss cone (of one or the other black hole) from the chaotic orbits induced by close interactions with the secondary. \citet{wegg11} and \citet{chen11} validated these findings and additionally accounted for the inspiral of the secondary, and also found that the importance of Kozai oscillations is \emph{reduced} by general relativistic precession and precession induced by the gravitational field of the cusp itself. \citet{lezhnin19} also recently confirmed this increase in the rate of TDEs, and accounted for the dynamical evolution of the distribution function of the stellar cluster, with Monte Carlo methods, {and \citet{li17} found a similar boost to the TDE rate by employing a direct, $N$-body integration of the inspiral phase.}  {Interestingly, even though this phase is very short-lived, the rate of TDEs can be so substantially increased that up to 10\% of all TDEs can be from this stage of the inspiral of the SMBH binary \citep{chen11}.}

As the secondary continues to inspiral through the cusp, the population of bound stars is depleted by three-body ejections and the binary continues to ``harden''{; these ejections occur primarily from the encounters between slowly-intruding (relative to the binary orbital speed) stars and the binary, and hence the separation at which the binary enters this hard phase is $a_{\rm h} \sim G\mu/\sigma^2$, where $\mu = M_1M_2/(M_1+M_2)$ (see also Equation \ref{ah}; \citealt{quinlan96}). For mass ratios of order unity, this separation is comparable to the sphere of influence of the pimary, but can be significantly smaller for very unequal-mass binaries}. \citet{chen08} studied the interactions between stars encountering a hardened binary, and showed that, because of the {the destruction of the stellar cusp during the hardening phase of the binary}, the rate of tidal disruptions can actually fall significantly below that of an isolated black hole. 

Finally, as the binary reaches semimajor axes such that the observed TDEs are expected to appear qualitatively differently from those by isolated SMBHs (semimajor axes of the order mpc and below; see Sections \ref{sec:elementary} and \ref{sec:fallback}), the speed of the binary greatly exceeds the velocity dispersion of the stars incident on the binary. In this case, the stars that encounter the binary have a binding energy of effectively zero at the large distances (relative to the binary separation) from which they originate, and the physical setup of the problem is similar to that of an isolated SMBH. However, the ``cross section'' of the binary is enhanced by a factor $a/r_{\rm t}$ (assuming that the stars entering the loss cone are in the pinhole regime), meaning that {-- for the same diffusion rate of stars into the loss cone --} the number of stars incident on the binary is larger {than that of an isolated black hole}. 

\citet{coughlin17} investigated the rate of tidal disruptions by a SMBH binary in this regime, focusing on binaries with a mass ratio between 0.1 and 1 (in increments of 0.1) and a separation of $a/r_{\rm t} = 100$; for a $10^{6} M_{\odot}$ primary, this separation is $a \simeq 0.1$ mpc. They found that the vast majority of stars encountering such a small-separation binary are ejected on hypervelocity orbits, with only $\sim$ 2\% being tidally disrupted by either the primary or the secondary; the total rate of disruption is also almost completely independent of the mass ratio (though the rate of disruption by the secondary scales almost exactly linearly with the mass ratio). However, the increased cross section of the binary implies that the rate of disruption \emph{relative to an isolated black hole} is $\dot{N}_{\rm TDE} = 2\%\times a/r_{\rm t} \simeq 2$. Thus, in this regime, the rate of disruption by the binary is augmented by a modest factor, which is due to the competition between the increased binary cross section and the prevalence of stellar ejection over disruption. \citet{darbha18} expanded upon the analysis of \citet{coughlin17} by studying the three-body interactions between incident, parabolic stars and binaries with a wide array of separations and mass ratios, and found that -- for the majority of mass ratios and separations -- the relative rate is augmented by a factor of $\sim$ 2-5. However, for certain combinations of binary properties that favor relatively wide separations and small mass ratios, the rate of disruption reaches a relative maximum that can be of the order 10.

In summary, binary black holes can generate a short-lived, $\lesssim 1$ Myr burst of TDEs as the secondary first encounters the cusp of stars around the primary {and reaches the hardening radius}. As the binary continues to harden, the rate of TDEs {can} drop below the single-black hole value {owing to the scouring out of the cusp and the potentially slow rate of repopulation of the loss cone}. Just prior to coalescence, when the binary reaches $\sim $ mpc separations, the {relative} rate {of tidal destruction} approaches and modestly exceeds (by a factor of 2-10) that of an isolated black hole.

\section{Fallback dynamics in black hole binaries}
\label{sec:fallback}
As we noted in Section \ref{sec:elementary}, there is expected to be a critical separation where the disrupted debris stream becomes significantly perturbed by the presence of the companion black hole on the orbital timescale of the binary. Within this separation, one expects the return rate of the debris to the disrupting hole to exhibit some modulation that is imparted by the non-Keplerian potential of the binary. As shown in Section \ref{sec:elementary}, this separation can be found by equating the Roche radius of the non-disrupting hole with the apocenter of the disrupted debris after the binary period $P$, which yields

\begin{equation}
    \frac{a}{r_{\rm t,1}} \simeq N(q)\left(\frac{GM_*}{4\pi^2R_*^3}\right)^{1/3}P^{2/3},
\end{equation}
where $N(q)$ is a numerical factor, of order unity, that depends weakly on the mass ratio (see Equation 2 of \citet{coughlin17} for the full equation). For periods on the order of months -- within which time we would have some chance of detecting the variability -- this expression yields separations on the order of hundreds of tidal radii.

After similarly motivating an appropriate selection of binary parameters, \citet{liu09} (see also \citealt{liu14}) used restricted three-body integrations to quantitatively assess the impact of the secondary on the fallback rate of debris to the primary (the disrupting black hole for their setup). They found that the fallback rate is indeed modified on timescales consistent with their analytic arguments, and for their specific setup (consisting of a $10^7 M_{\odot}$ primary) it exhibited dramatic ``dips'' around five to ten years. These dips were displayed when the secondary came close enough to the disrupted debris stream to significantly alter its dynamics, thereby deflecting it from its return path to the primary.

Using similar numerical techniques, \citet{ricarte16} extended the work of \citet{liu09} by performing a more expansive parameter space study, and investigated the response of the fallback rate -- and in particular the presence and longevity of the interruptions -- on the binary mass ratio, the primary mass, and the binary separation. They also accounted for the fallback rate onto the secondary. While certain binary parameters exhibited unambiguous dips in the total accretion rate (primary plus secondary), in others, and in particular for more equal mass ratios, the reduction in the fallback rate of the primary was supplemented by a nearly equal increase in the fallback onto the secondary. 

While these first pioneering studies validated the basic intuition regarding the impact of the secondary on the fallback, a number of aspects of the fallback remained unanswered. In particular, stream self-intersections should, non-trivially, influence and likely augment the accretion rate onto both black holes, as suggested by \citet{ricarte16}, if the hydrodynamics of the gas is taken into account. Likewise, at small scales ($\lesssim r_{\rm t}$) around the black holes the debris should circularize viscously into discs; therefore, the abrupt and complete shutoff of the fallback of debris found by \citet{chen09} should be less pronounced if the inner regions around the black holes are resolved and the material is allowed to self-interact. Owing to its reduced Hill sphere, disruptions by the secondary, while statistically less likely \citep{chen08, chen09}, should also exhibit a greater degree of irregularity in the fallback rate and debris morphology.

Additionally, \citet{liu09} and \citet{ricarte16} assumed that the frozen-in approximation was upheld by the disrupted star prior to reaching pericenter, and thereby let the stellar center of mass be described by a parabolic orbit up until the moment at which the star entered the tidal sphere around the disrupting SMBH. However, the star can interact strongly with, and enter into a long-lived, chaotic orbit about, the binary for many orbital periods \emph{prior} to disruption. In these instances, the specific energy and angular momentum of the stellar COM could be modified significantly from their fiducial values (being 0 and $\sqrt{2GMr_{\rm p}}$ respectively, where $r_{\rm p}$ is the pericenter distance of the stellar center of mass to the disrupting black hole of mass $M$), which would correspondingly introduce variations in the accretion rates. 

\citet{coughlin17} performed a study of both the three-body (i.e., stellar COM specific energy and angular momentum-altering) and hydrodynamic effects of the binary on the fallback of tidally-disrupted debris to both the SMBHs. By performing millions of restricted three-body integrations by binaries with mass ratio in the range 0.1 -- 1 (in increments of 0.1) and a separation $a/r_{\rm t,1} = 100$, they found that the stellar specific energy could be significantly modified from its original, parabolic value. The distribution of $\beta = r_{\rm t}/r_{\rm p}$ for disrupted stars was found to be very well reproduced by a $1/\beta^2$ falloff, indicating that the angular momentum distribution of disrupted stars is roughly uniform. Some ($\gtrsim 10\%$) tidally disrupted stars were also found to have one or more ``close encounters'' with one or the other black hole, defined to be instances in which the star came within three tidal radii of either hole. In these cases, partial disruptions could occur or significant rotation could be induced to the star prior to its full disruption (for the impact of stellar rotation on the disruption process see \citealt{golightly19}).

To investigate the effects of hydrodynamics on the evolution of the tidally-disrupted debris, \citet{coughlin17} randomly selected 120 of the three-body encounters that resulted in the disruption of the star, and used their orbital parameters to initialize smoothed-particle hydrodynamics (SPH) simulations (with the code {\sc phantom}; \citealt{price17}) of the disruption of a Solar-like star by a binary with primary mass $M_1 = 10^6M_{\odot}$, secondary mass $M_2 = 2\times 10^{5}M_{\odot}$, and separation $a = 100\, r_{\rm t 1} \simeq 0.2$ mpc. They found that the modifications to the specific energy of the center of mass could have significant effects on the return of the debris, and these effects were isolated from the post-disruption influence of the companion black hole by performing ``control'' simulations; in the controls, the non-disrupting black hole was removed, but the same initial conditions (in the rest frame of the disrupting hole) were kept, thereby singling out the contributions to the fallback from the change in COM properties. In extreme cases, all of which were disruptions by the secondary, the stream was completely ejected from the binary and resulted in no accretion onto either black hole owing to the large (positive) shift in the COM energy. In other instances where the stellar COM was placed on a bound orbit prior to disruption, the return of the most-bound debris could be significantly shorter than the parabolic value, with an appropriately increased peak fallback rate. 

In general agreement with \citet{liu09} and \citet{ricarte16}, the accretion rates measured by \citet{coughlin17} exhibited dips after a characteristic timescale comparable to a fraction of an orbital period (being $T_{\rm orb} \simeq 106$ d). However, the relative minima attained by these dips only amounted to fractional reductions of 1 to 10\%, due to the fact that small-scale accretion disks around the individual black holes continued to supply them with a finite amount of gas. Moreover, there were some rarer instances in which the accretion rate actually \emph{increased} relative to the control simulation, and these cases tended to occur when the returning debris stream, the disrupting black hole, and the binary center of mass were all roughly colinear (see also \citealt{coughlin18}). 

\begin{figure}
    \centering
    \includegraphics[width=1\textwidth]{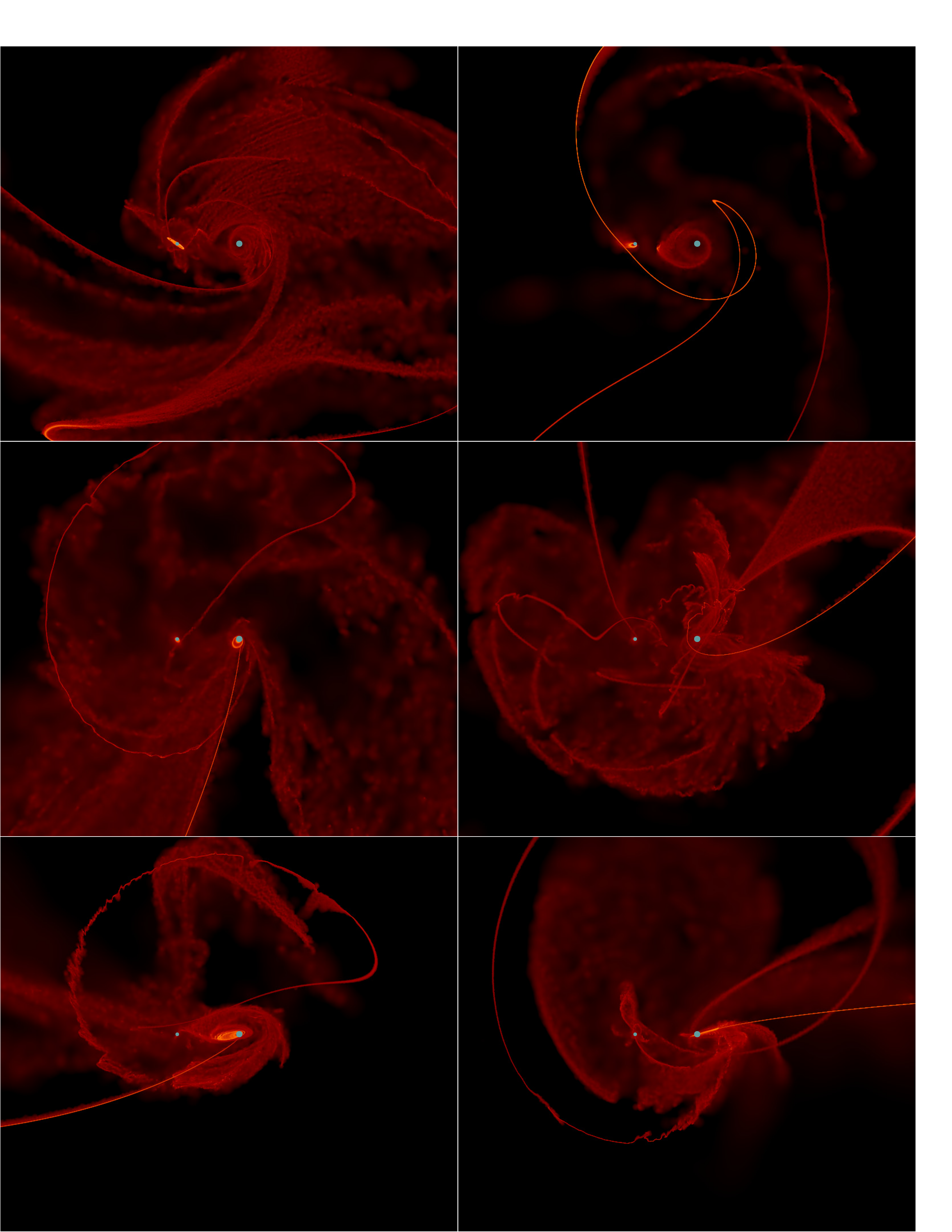}
    \caption{The morphology of the tidally-disrupted debris resulting from six different tidal disruptions of stars by a supermassive black hole binary (adapted from \citealt{coughlin17}); bright (dark) colors indicate regions of increased (reduced) {column} density {integrated normal to} the plane of the binary, and the blue circles indicate the supermassive black holes. It can be seen that accretion onto one or the other black hole can occur, and there are a number of ``gaps'' where the stream self-intersects. The debris also expands to very large radii and angles and surrounds the SMBH binary.}
    \label{fig:multiple_TDEs}
\end{figure}

In some instances, the morphology of the tidally-disrupted debris resembled a ``cloud'' that enshrouded the binary itself, and extended out to a large number of binary semimajor axes; there were also a modest number of disruptions -- especially those that were ejected nearly perpendicular to the binary orbital plane -- that were far more ``tame'' in appearance, with debris primarily accreting onto one or the other black hole in a manner roughly consistent with the isolated black hole picture. In line with the expectations of \citet{ricarte16}, a large number of self-intersections occurred when the stream was confined to within a small angle of the binary orbital plane, which likely contributed an additional source of energy dissipation and enhanced the accretion rates of the holes. Figure \ref{fig:multiple_TDEs} shows six examples of the morphology of the disrupted debris that resulted from the simulations in \citet{coughlin17}. 

\begin{figure*}
\centering
\includegraphics[width=0.9\textwidth]{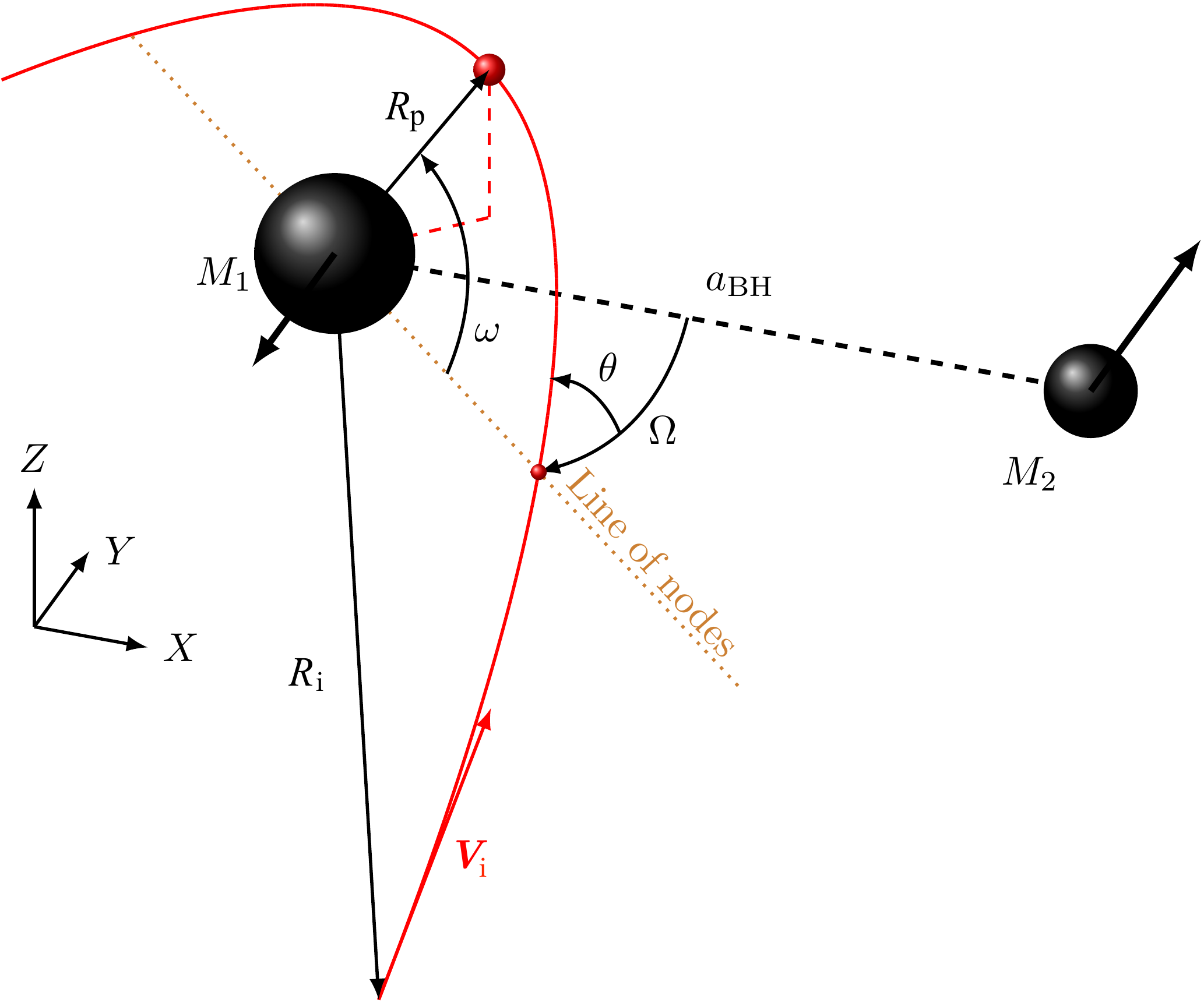}
\caption{Schematic of the geometry of the disruption of a star by a binary black hole. From \citet{vigneron18}.}
\label{fig:scheme_vigneron}
\end{figure*}

\begin{figure*}
\centering
\includegraphics[width=1\textwidth]{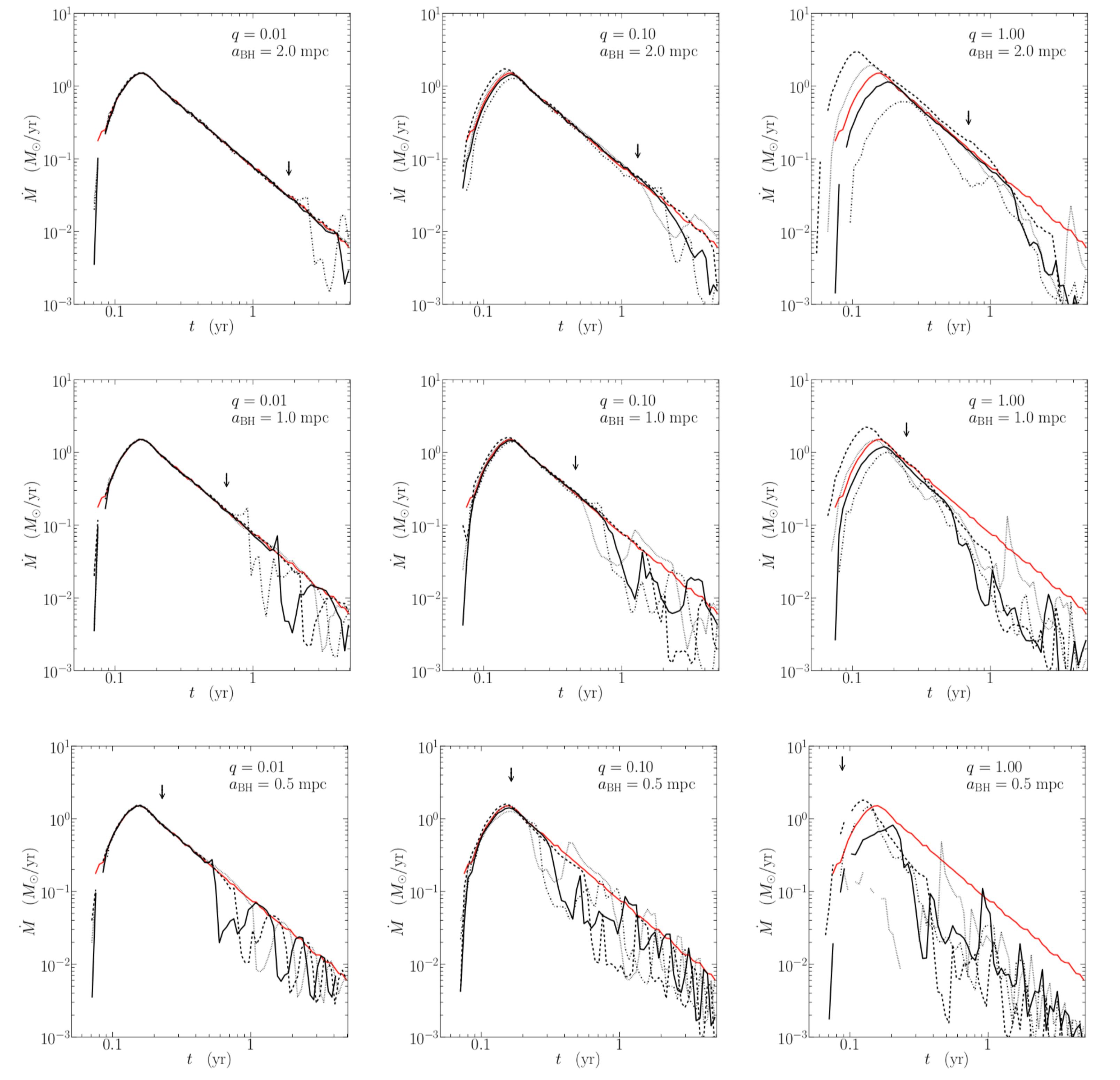}
\caption{Fallback rate of TDE from a binary black hole system. The red curve shows the ``control'' run of a TDE from a signle black hole, while the various lines refer to different values of the angle $\Omega$. In all these simulations the stellar orbit was in the same plane as the binary orbit. See \citet{vigneron18} for details.}
\label{fig:vigneron1}
\end{figure*}

\citet{vigneron18} performed a more controlled study of the effects of several physical and geometrical parameters involved in the disruption by a binary black hole, with the aim to use TDE fallback rates as probes of hidden sub-pc black hole binaries. The main binary parameters are the mass of the primary black hole (held constant to $10^6M_{\odot}$ by \citealt{vigneron18}), the mass ratio $q$, the binary separation $a$ and the eccentricity of the binary orbit (assumed to be equal to zero). The stellar orbit is characterized by its eccentricity, assumed to be equal to 1 (i.e. a parabolic orbit), its pericenter distance (assumed to be equal to the tidal radius with respect to the primary black hole) and the three angles shown in Fig. \ref{fig:scheme_vigneron}, indicating the inclination of the stellar orbit with respect to the binary orbit $\theta$, the angle of the line of the nodes $\Omega$ and the apsidal angle of the pericenter $\omega$. \citet{vigneron18} performed simulations by varying the mass ratio, the binary separation and the angles $\theta$, $\Omega$ and $\omega$. Only disruptions from the primary black hole were considered. Firstly, it is important to define two fundamental scales for the binary separation. For very close binaries, the apocenter of the most bound debris after disruption will be larger {than} the Roche lobe of the primary. In this case, all the debris will be perturbed by the binary and the fallback rate will not show a power-law decay at any time. Conversely, for large enough binary separation, the debris will be affected by the presence of the binary only at very late times, when the fallback is unlikely to be observable. If we assume that this happens when the fallback rate has reached 1\% of its peak value, the two critical binary separations are given by:
\begin{eqnarray}
		a_{\rm min} = \dfrac{0.6q^{2/3}+\ln{\left(1+q^{1/3}\right)}}{0.49q^{2/3}} R_\star \left(\dfrac{M_1}{M_\star}\right)^{2/3}, \\
		a_{\rm max} = \epsilon^{-2/5} a_{\rm min}
	\label{eq::a_minmax}
\end{eqnarray}
where $\epsilon=0.01$. For typical parameters, $a_{\rm min}\approx 0.4-0.6$ mpc and $a_{\rm max}\approx 2-4$ mpc. For binary separations in between the two critical values, we thus expect a TDE to initially show the standard $t^{-5/3}$ fallback curve, but to then experience dips and interruptions as found by previous studies. The numerical simulations were performed using the same {\sc phantom} code \citep{price17} as \citet{coughlin17}. The results confirm the expectations, in that for separations smaller than $a_{\rm min}$ the fallback curve shows a series of quasi-periodic dips (with a period of the order of the binary period) and no immediately recognizable power-law decay. On the contrary, for separations of the order of $a_{\rm max}$ the fallback proceeds undisturbed, and follows the trend of a TDE by a single black hole. Figure \ref{fig:vigneron1} shows the fallback rates of several simulations, where $\theta=0$ (disruption in the binary plane). The top left panel refers to a separation that is just above $a_{\rm max}$, while the bottom right is just within $a_{\rm min}$. The red curve refers to a TDE from a single black hole, and the various curves refer to different choices of $\Omega$. 

\begin{figure*}
\centering
\includegraphics[width=1\textwidth]{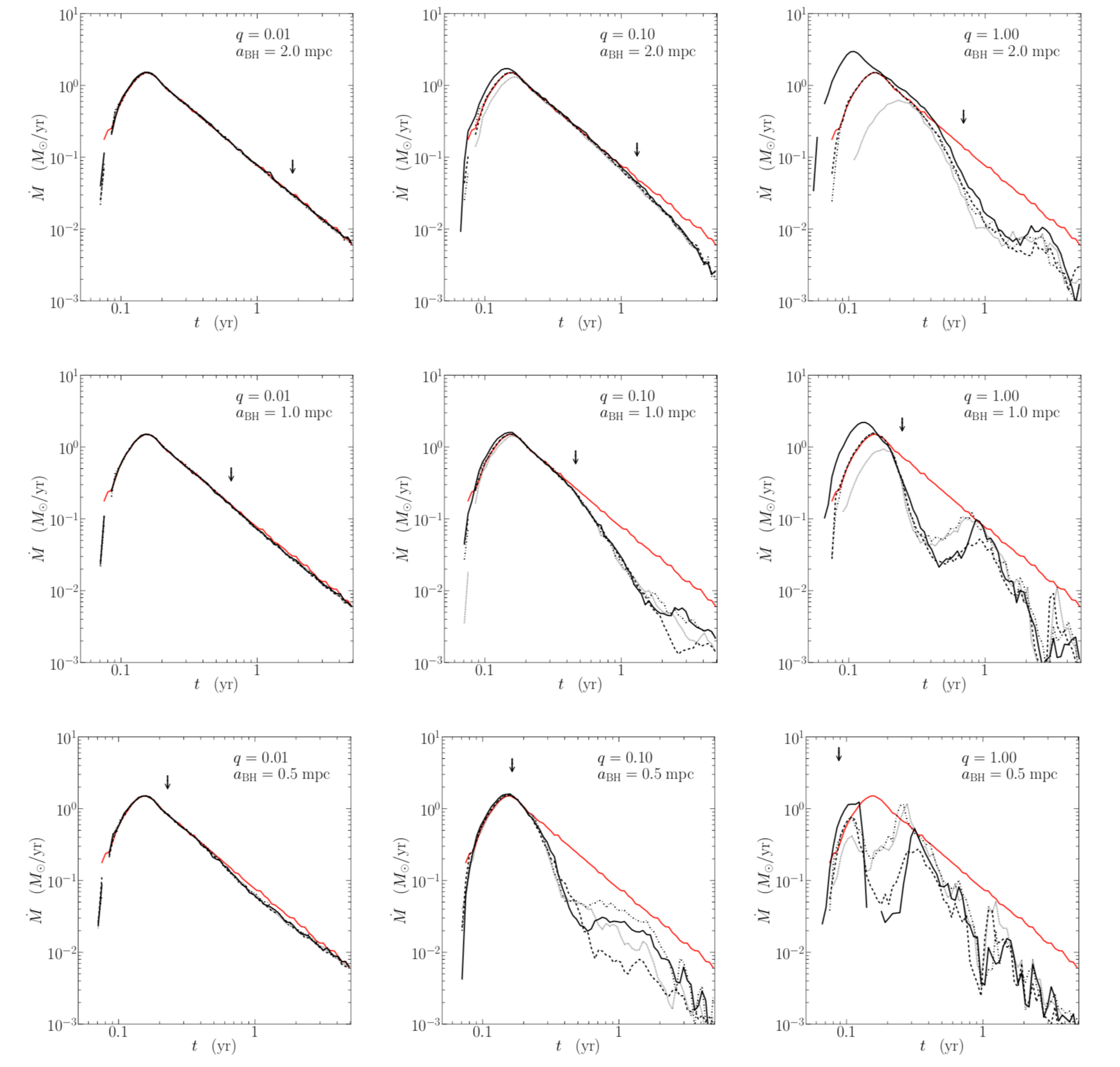}
\caption{Same as Figure \ref{fig:vigneron1}, but for a polar stellar orbit. See \citet{vigneron18} for details.}
\label{fig:vigneron2}
\end{figure*}

An interesting behaviour is seen when comparing disruptions in the plane of the binary (that is, with $\theta=0$) to disruptions perpendicular to the binary orbit ($\theta=\pi/2$). For perpendicular stellar orbits, rather than a series of periodic dips, the fallback rate shows a smooth decline below the expected, $t^{-5/3}$ curve, with a possible recovery only at late times and again in a relatively smooth fashion, with respect to the sudden interruptions seen in the in-plane disruption. These are shown in Figure \ref{fig:vigneron2}, which shows the same fallback rates as Figure \ref{fig:vigneron1}, but for a polar stellar orbit. Such smooth decline and re-brightening of the fallback luminosity has been observed in the lightcurve of ASASSN-15lh \citep{leloudas16}, which has indeed been suggested to be a TDE from a binary black hole \citep{coughlin18}.

Finally, we note that while we focused on the case of fallback dynamics generated by tidal disruption events generated by supermassive black hole binaries, there has also been interest in the effects of tidal disruption events by \emph{stellar mass} black holes in binary systems. \citep{lopez18} investigated the hydrodynamics of these events numerically, and found that such disruptions can potentially alter the mass and spin distributions of stellar mass black holes. \citet{samsing19} also showed that the electromagnetic signatures of TDEs by stellar mass black hole binaries in globular clusters could probe the evolution of the binary fraction, and the orbital properties of the binaries themselves, within those clusters.

\section{Prospects for detection}
\label{sec:prospects}
From Section \ref{sec:rates}, the most pronounced effects of the binary in terms of the rates of disruption occur when the secondary first encounters the cusp of stars around the primary. For typical primary masses, this initial, burst of TDEs will occur {as} the binary {spirals in to the hardening radius, which is of the order} $\sim 0.01 - 10$ pc separations {depending on the black hole mass ratio and the stellar velocity dispersion (see Equation \ref{ah})}. Unfortunately, since the tidal disruption radius of either black hole is on the order of tens of gravitational radii, such separations are millions of tidal radii. Therefore, even though one could detect more than one tidal disruption event from the same galaxy during this phase (over the course of a human lifetime, especially during the upcoming age of LSST; see Table 1 in \citealt{wegg11}), {which would be good indirect evidence for the existence of a black hole binary,} the influence of the binary companion on the fallback of debris -- which would enable the more direct detection of the second black hole -- will be completely negligible. 

As noted in Sections \ref{sec:elementary} and \ref{sec:fallback}, the binary companion is expected to, and has been verified numerically to, modify the accretion process onto the disrupting black hole when the fallback time of the most-bound debris is comparable to the binary orbit, and typical black hole masses then yield a requisite binary separation on the order of $\sim$ mpc. At these separations, the binary has entered the gravitational-wave inspiral regime (i.e., the emission of gravitational waves will cause the binary to merge within the Hubble time), and the time to merger is \citep{peters64}

\begin{multline}
T_{\rm GW} = \frac{5}{256}\frac{c^5}{G^3}\frac{a^4}{M_1M_2\left(M_1+M_2\right)}  \\ \simeq 5 \left(\frac{a}{0.23\text{ mpc}}\right)^4\left(1+\left(\frac{q}{0.2}\right)^{-1}\right)\left(\frac{M}{10^6M_{\odot}}\right)^{-3} \text{ Myr}
\end{multline}
(Note that $0.23$ mpc is roughly 100 times the tidal radius of a $10^6M_{\odot}$ SMBH). Thus, by the time the binary reaches the separations at which TDEs are expected to appear distinct from single-SMBH disruptiions, the lifetime of the binary is cosmologically short. Therefore, the vast majority of TDEs that are caused by binary black holes will occur when the binary separation is too wide to permit direct detection (though, as noted above, multiple TDEs in the same galaxy would be very good, indirect evidence for the existence of a SMBH binary, and {it is predicted} that LSST {may} observe such multiple-TDE galaxies {at a rate comparable to one per year}; \citealt{wegg11,thorp18}).

Nevertheless, there have been suggestions that some observed transients originate from tidal disruptions by black holes in small-separation binaries. For example, the lightcurve from the galaxy SDSS J1201+30, which was previously classified as a candidate TDE, exhibited dramatic reductions in luminosity over a timescale of years. From the results of \citet{liu09}, \citet{liu14} argued that these sudden dips could indicate a binary companion at a separation of $\sim$ 1 mpc.  However, the sparse coverage of the source does not rule out variability on smaller timescales, which could be due to inherent AGN activity (e.g., \citealt{grupe15}). 

It has also been suggested that the event ASASSN-15lh, originally described by \citet{dong16} as one of the most luminous supernovae ever observed, was caused by the disruption of a star by a black hole in a binary system. In particular, its puzzling, ``double-humped'' nature in the UV can be explained by a secular change in the accretion rate onto the disrupting black hole, that change induced by the perturbing gravitational presence of a second black hole. If one requires that the time to rebrightening of $\sim 100$ days equal the time for the disrupted debris to reach the Hill sphere of the primary, the mass of which being $\sim 10^{8}M_{\odot}$ from the $M$-$\sigma$ relation of the host galaxy, then one finds that a secondary black hole with mass $\sim 5\times 10^{5} M_{\odot}$ at a separation of $\sim 0.5$ mpc can reproduce the observed variation (and was reproduced with the simulations of \citealt{coughlin18}). However, there have also been other interpretations of this event, including the tidal disruption of a star by a rapidly-rotating black hole \citep{leloudas16}, fallback onto a newly-formed magnetar \citep{moriya16,metzger18,margalit18}, and the sequential, tidal disruption of two stars previously in a stellar binary \citep{coughlin18b}. 

While the likelihood of there being a tidal disruption in the final Myr of the existence of a supermassive black hole binary may be relatively low, there are other features of such tidal disruption events that may make them easier to detect. For one, there is an additional, apsidal precession of the returning debris stream that is caused by the gravitational influence of the second black hole, and this deflection -- which can exceed the general relativistic one -- creates more efficient dissipation of the kinetic energy of the debris. Indeed, this self-intersection was observed in the simulations of \citet{coughlin17}, in spite of the fact that general relativity was not included. As a consequence, disc formation and the brightening of the source may be less adversely affected by Lense-Thirring precession, which torques the stream out of the orbital plane and causes the two streams to miss one another \citep{hayasaki13}, and the ``dark years'' that may plague low-$\beta$ disruptions \citep{guillochon15} can be more easily avoided. 

In addition, there are often multiple, distinct accretion episodes that occur from the disruption of a star by a black hole in a binary, as both black holes can actively accrete the stellar debris and the gravitational field of the second black hole induces variation in the binding energy of the disrupted material. Furthermore, the timescales associated with the brightenings of each flare are not necessarily limited by the intrinsic fallback time to the disrupting black hole, as shown in the simulations of \citet{coughlin17}, and the luminosities can -- in some instances -- exceed the one predicted for an isolated SMBH. Thus, while a single flare may be too dim or fade too rapidly for the cadence of current and even upcoming surveys, the recurrent and larger flares induced by the second hole can be more readily detected. 

As noted in Section \ref{sec:fallback}, the pre-disruption orbit of a star in the binary potential can also induce large changes in the orbital properties of the center of mass of the star. As shown by \citet{darbha18}, the greatest changes to the binding energy occur when the binary is at its smallest separations, and in the most extreme cases can result in completely bound debris streams. In these scenarios, the accretion rate is greatly increased above the standard, isolated-black hole rate, which makes such systems much more luminous. In other instances the center of mass receives a large, positive-energy kick, which can result in completely \emph{unbound streams}. While there is no accretion luminosity from such TDEs, the radio emission from the unbound ejecta (as it shocks against the circumnuclear medium; \citealt{guillochon16, yalinewich19}) will be more luminous owing to its increased kinetic energy budget, which may also enhance the detectability (albeit outside the context of optical surveys).

Finally, even if the tidal disruption event from a small-separation, SMBH binary is not detected, the tidally-disrupted debris can provide an additional reservoir of gas in the vicinity of the binary at the time of merger. As shown by \citet{coughlin17b}, circumbinary rings can form out of the debris streams for binaries with small mass ratios, which then spread viscously into a disc over many binary orbits. This disc of gas then responds dynamically to the mass lost to gravitational waves at the time of merger, the electromagnetic signature from which being potentially detectable in nearby galaxies \citep{rossi10,corrales10,rosotti12}

\section{Summary}
\label{sec:summary}
In this chapter we presented an overview of the current understanding of the influence of black hole binarity on the properties of tidal disruption events. As described in Section \ref{sec:elementary}, the presence of a second black hole can affect both the rate at which TDEs occur and the evolution of the fallback. The magnitude of each effect is, however, highly-dependent on the separation of the binary.

Past work on the rates of TDEs by black holes in binary systems (Section \ref{sec:rates}) has indicated that, {as the secondary spirals in through the cusp of the primary and approaches the hardening radius,} the rate can be appreciably augmented (by 2-3 orders of magnitude above the isolated-black hole rate) owing to Kozai oscillations and direct interactions between the secondary and bound stars. As the binary continues to harden, the depletion of the population of bound stars {and the potentially-slow rate of refilling of the loss cone can} cause the TDE rate to plummet below the single-black hole case. Once the binary reaches very small separations, such that the binary orbital velocity greatly exceeds the stellar velocity dispersion, the rate increases slightly above (by a factor of a few to 10 at most) that of an isolated black hole {with the same rate of refilling of the loss cone}.

While the TDE rate is modified once the binary separation is comparable to the sphere of influence of the primary, being $\sim$ pc scales for typical galaxy types and black hole masses, the fallback of debris onto a black hole in a binary system (see Section \ref{sec:fallback}) is only dramatically altered once the binary shrinks to $\sim$ mpc separations. As argued in Section \ref{sec:elementary}, at these small separations the binary orbital time is comparable to the fallback time of the most bound debris, meaning that the dynamics of the returning gas stream will be affected by the secondary in the early stages of the accretion process. This basic argument has been substantiated with numerical simulations that investigated both the gravitational and hydrodynamical evolution of TDEs from binary SMBHs.

Because of the small separations required to alter the fallback of tidally-disrupted debris, SMBH binaries that yield directly observable deviations from the single-black hole predictions are well into the gravitational-wave inspiral regime and are correspondingly short-lived; the overall rate of detection of TDEs by binary SMBHs (Section \ref{sec:prospects}) is therefore reduced by this relatively narrow window during which we can observe such binaries. However, the features of the TDEs themselves, including much greater variability and increased accretion rates (and likely luminosities), likely render them easier to detect over single-SMBH TDEs, and there have been some claims of detections. Moreover, the extremely high rate of TDEs by binaries at larger separations implies that multiple TDEs may be discovered in the same galaxy, especially in the current era of wide-field optical surveys. While observing multiple TDEs from the same galaxy would (likely) not yield direct evidence in the form of variation in the accretion luminosity, it would provide good, indirect evidence of the existence of SMBH binaries.

\section*{Acknowledgments}
ERC acknowledges support from NASA through the Einstein Fellowship Program, grant PF6-170170, and the Hubble Fellowship, grant \#HST-HF2-51433.001-A awarded by the
Space Telescope Science Institute, which is operated by
the Association of Universities for Research in Astronomy, Incorporated, under NASA contract NAS5-26555. PJA acknowledges support from NASA through grant NNX16AI40G. CJN is supported by the Science and Technology Facilities Council (STFC) (grant number ST/M005917/1)

\bibliographystyle{aps-nameyear}
\bibliography{refs}

\end{document}